\documentclass[twocolumn]{IEEEtran}  
\IEEEoverridecommandlockouts
\makeatletter
\def\ps@IEEEtitlepagestyle{%
  \def\@oddfoot{}%
  \def\@evenfoot{}%
  \def\@oddhead{}%
  \def\@evenhead{}%
}
\makeatother
\usepackage{nicematrix}
\usepackage{cite}
\usepackage{amsmath,amssymb,amsfonts}
\usepackage{mathtools, nccmath}
\usepackage{algorithmic}
\usepackage{graphicx}
\usepackage{textcomp}
\usepackage{xcolor}
\usepackage{diagbox}
\usepackage{float} 
\def\BibTeX{{\rm B\kern-.05em{\sc i\kern-.025em b}\kern-.08em
    T\kern-.1667em\lower.7ex\hbox{E}\kern-.125emX}}
\usepackage{setspace}
\begin{document}
\title{Impact of Reflectors and MIMO on ML-Aided mmWave/sub-THz Blockage Prediction\\
\thanks{This work was partially supported by National Science Foundation Grant
ECCS-2122012.}
}

\author{\IEEEauthorblockN{Roghieh Mahdavihaji\textsuperscript{$\dagger$}, Alexandra Duel-Hallen\textsuperscript{$\dagger$}, Hans Hallen\textsuperscript{$\ddag$}}
\IEEEauthorblockA{\textit{Department of Electrical and Computer Engineering, North Carolina State University, Raleigh, NC, USA\textsuperscript{$\dagger$}} \\
\textit{Physics
Department, North Carolina State University, Raleigh, NC, USA\textsuperscript{$\ddag$}}\\
Emails:{\{rmahdav2, sasha, hallen\}}@ncsu.edu}
}
\maketitle
\thispagestyle{empty}
\pagestyle{empty}

\begin{abstract}
The performance of millimeter-wave (mmWave) and sub-terahertz (sub-THz) communication systems is significantly impaired by sensitivity to sudden blockages. In this work, we employ machine learning (ML) and our physics-based simulation tool to warn about the upcoming blockage tens of 5G frames ahead for highway speeds, providing a sufficient time for a proactive response. Performance of this ML-aided early-warning-of-blockage (ML-EW) algorithm is analyzed for realistic outdoor mobile environments with diverse reflectors and antenna arrays placed at the base station (BS) and user equipment (UE) over a range of mmWave and sub-THz frequencies. ML accuracy of about 90\% or higher is demonstrated for highway UE, blocker, and reflector speeds,  multiple-input-multiple-output (MIMO) systems, and frequencies in the mmWave/sub-THz range.

\end{abstract}


\section{INTRODUCTION}
Utilization of millimeter wave (mmWave) and sub-THz frequencies  can potentially boost the  throughput of wireless communication systems since these frequencies provide much wider available bandwidth than the  sub-6 GHz frequencies. However, there are inherent challenges to effective deployment of mmWave/sub-THz technology. In particular, line-of-sight (LoS) blockage by static or moving physical objects can severely limit performance of these systems \cite{b2} due to more abrupt variation of the received signal strength (RSS) caused by weaker diffraction and greater penetration loss than for sub-6GHz frequencies \cite{b3}.

In \cite{b8}, we employed the MiniRocket Machine Learning (ML) time-series classifier method \cite{b4} to provide early warning (EW) of LoS mmWave blockage hundreds of ms ahead for highway speeds using in-band signal observations, thus enabling a proactive response to an upcoming blockage, e.g., a base station (BS) handover, beam switching, or modified resource allocation. The mmWave signal datasets for training and testing the ML method \cite{b8} were created using our physics-based simulation tool, which models reflection and diffraction accurately \cite{b7,b8}. Our insights and numerical results suggest that the early warning of blockage is possible due to diffraction-induced pre-blockage signal patterns. However, \cite{b8} did not investigate the impacts of reflectors with varying sizes and angles, multiple-input-multiple-output (MIMO), and various carrier frequencies on the EW performance. In this work, we include the latter features into the ML training and testing data sets for multiple mmWave and sub-THz frequencies and validate the ML-aided EW of blockage capability in these realistic mobile wireless scenarios.

Several other works employed pre-blockage channel characteristics for temporal prediction of mobile mmWave blockage. In \cite{b7}, LoS blockage was {forecast based on sub-6 GHz observations using our realistic physical model. Fresnel-Kirchhoff diffraction modeling was employed in \cite{b9} to detect amplitude fluctuations and diffraction fringe characteristics of low-frequency signals, which were used to warn about upcoming blockage of mmWave signals. In \cite{b12}, in-band mmWave measurements were employed in recursive neural network (RNN) and convolutional neural network (CNN) {models for predicting upcoming LoS blockage using wireless signatures.

Unlike \cite{b9,b12} and most other prior works on mmWave blockage prediction, this paper does not assume specific topology, directions, or speeds. Moreover, our approach \textit{relies solely on in-band mmWave or sub-THz signal observations} and does not require sub-6 GHz observation or cameras. Furthermore, \textit{we employ the MiniRocket} \cite{b4} \textit{method}  that is \textit{less complex} and \textit{much faster} than deep learning (DL) models. Finally, \textit{we investigate the impacts of reflectors in the environment, MIMO configurations, and sub-THz frequencies on the EW of LoS blockage, which have not been studied previously}.   

The rest of the paper is organized as follows. The physics-based channel model is summarized in Section II,  highlighting the oscillation patterns associated with LoS blockages and reflectors. The ML method for early warning of blockage is presented in Section III. Section IV contains simulation results for the ML-EW algorithm and discusses the impacts of reflectors in the environment, MIMO and the carrier frequency on the RSS and EW accuracy. Finally, conclusions are presented in Section V.

\section{CHANNEL MODEL}

The physical model used in this paper is based on the method of images and Fresnel diffraction as detailed in \cite{b7,b8, b18, b19}. As in \cite{b8}, we employ this model to simulate diverse blockage and non-blockage scenarios to create the training and testing datasets for the ML-EW method. However, in contrast to \cite{b8}, this paper considers the presence of both static and moving reflectors with varying sizes and multiple antennas. Fig. \ref{fig1} depicts an example of the physical model scenario with stationary BS at (${x_{bs}}$, ${y_{bs}}$) and blocker (e.g., a parked bus) at (${x_{b}}$, ${y_{b}}$) implemented as an edge for diffraction in the physical model \cite{b7}. The UE is initially positioned at (${x_{0}}$,${y_{0}}$) in the LoS region and then moves with constant speed $v$ on a straight line to (${x_{1}}$,${y_{1}}$) in the NLoS region. Moreover, Fig. \ref{fig1} shows $Q = 4$ significant reflectors affecting the received signals as detailed below.  

\begin{figure}[]
\centerline {\includegraphics[width=0.505\textwidth, height=9.8cm]{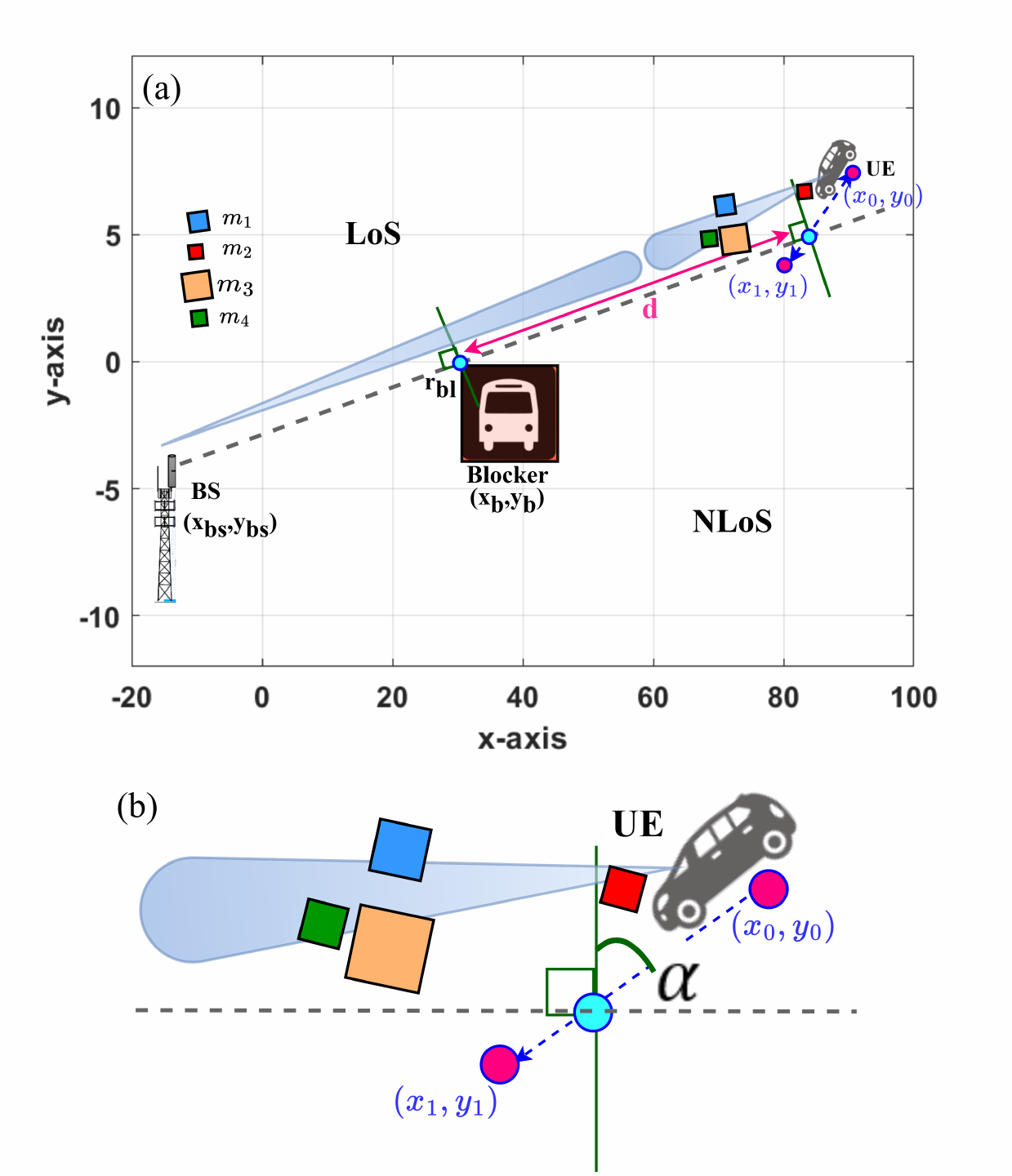}}
\caption{Example simulation scenario for the physical model; (a) scenario with $Q$ = 4 significant reflectors; (b) zoom-in of upper right corner of (a) and rotated so LoS/NLoS edge (dashed line) is horizontal rather than x-axis in (a).}
\label{fig1}
\end{figure}
\setlength{\textfloatsep}{6pt plus 1.3 pt minus 7.0pt}

We employ uniform linear arrays (ULA) with $L$ antennas in BS and $M$ antennas in UE. Assume carrier frequency $f_{k}$. The phase shifts that must be applied to the transmit and receive antennas in order to point the beams toward azimuth angles $\theta_{t}$ and $\theta_{r}$ relative to the endfire directions, the axes of the arrays, are given by the corresponding steering vectors \cite{b15} 

\begin{equation}
\textbf{a}_{t}(\theta_{t}) = [1 \quad e^{j\frac{2\pi\Delta_t\cos{\theta_{t}}}{\lambda_k}} ... \quad e^{j\frac{2\pi(L-1)\Delta_t\cos{\theta_{t}}}{\lambda_k}}]^T ,
\label{eq1}
\end{equation}

\begin{equation}
\textbf{a}_{r}(\theta_{r}) = [1 \quad e^{j\frac{2\pi\Delta_r\cos{\theta_{r}}}{\lambda_k}} ... \quad e^{j\frac{2\pi(M-1)\Delta_r\cos{\theta_{r}}}{\lambda_k}}]^T,
\label{eq2}
\end{equation}

\noindent where $^T$ is the transpose operator, $\lambda_k$ is the wavelength for frequency $f_k$ and $\Delta_{t}$ and $\Delta_{r}$ are the distances between two adjacent antenna elements of the transmitter (Tx) and receiver (Rx) antennas, respectively. Suppose there are $Q$ \textit{significant} reflectors ($q =1, . . . ,Q$), i.e., the reflectors within main or side lobes of the Tx and Rx antennas for at least some portion of the UE trajectory. Assuming narrowband transmission (flat fading), the equivalent lowpass gain of the channel from the first element of the transmitter antenna to the first element of receiver antenna, which correspond to the first elements of the steering vectors (\ref{eq1}, \ref{eq2}), is given by $h_{k,q}(t)$ for the $q^{th}$ multipath component (MPC).  The 'direct', or non-reflected, MPC is represented by the $q = 0$ value, $h_{k,0}(t)$. In these frequency-dependent gains, we account for the diffraction effects as described in eq. (2) of \cite{b7}. The equivalent lowpass gain of the channel at frequency $f_{k}$ that incorporates antenna arrays is given by

\begin{equation}
h'_{k}(t) = \textbf{V}^H_{r}(\theta_{r,q}) \left( \sum_{q=0}^{Q} h_{k,q}(t) \;  \mathcal{\textbf{a}}_{r}(\theta_{r,q}) \;\mathcal{\textbf{a}}^H_{t}(\theta_{t,q}) \right) \textbf{V}_{t}(\theta_{t,q}),
\label{eq3}
\end{equation}

\noindent where $^H$ is the Hermitian (conjugate transpose) operator and $\theta_{t,q}$ and $\theta_{r,q}$ are the angles of departure and arrival of the $q^{th}$ path. In (\ref{eq3}), we assume that the antennas are optimally aligned for the LoS signal, so the $L\times 1$ and $M\times 1$ unit-magnitude beamforming vectors at BS and UE, respectively, are computed as \cite{b15} 

\begin{equation}
\textbf{V}_{t}(\theta_{t,0}) = \frac{1}{\sqrt{L}} [1 \quad e^{j\frac{2\pi\Delta_t\cos{\theta_{t,0}}}{\lambda_k}} ... \quad e^{j\frac{2\pi(L-1)\Delta_t\cos{\theta_{t,0}}}{\lambda_k}}]^T,
\label{eq4}
\end{equation}

\begin{equation}
\textbf{V}_{r}(\theta_{r,0}) = \frac{1}{\sqrt{M}} [1 \quad e^{j\frac{2\pi\Delta_t\cos{\theta_{r,0}}}{\lambda_k}} ... \quad e^{j\frac{2\pi(M-1)\Delta_t\cos{\theta_{r,0}}}{\lambda_k}}]^T.
\label{eq5}
\end{equation}

\begin{figure}[]
\centerline{ \includegraphics[width=0.52\textwidth, height=9.cm]{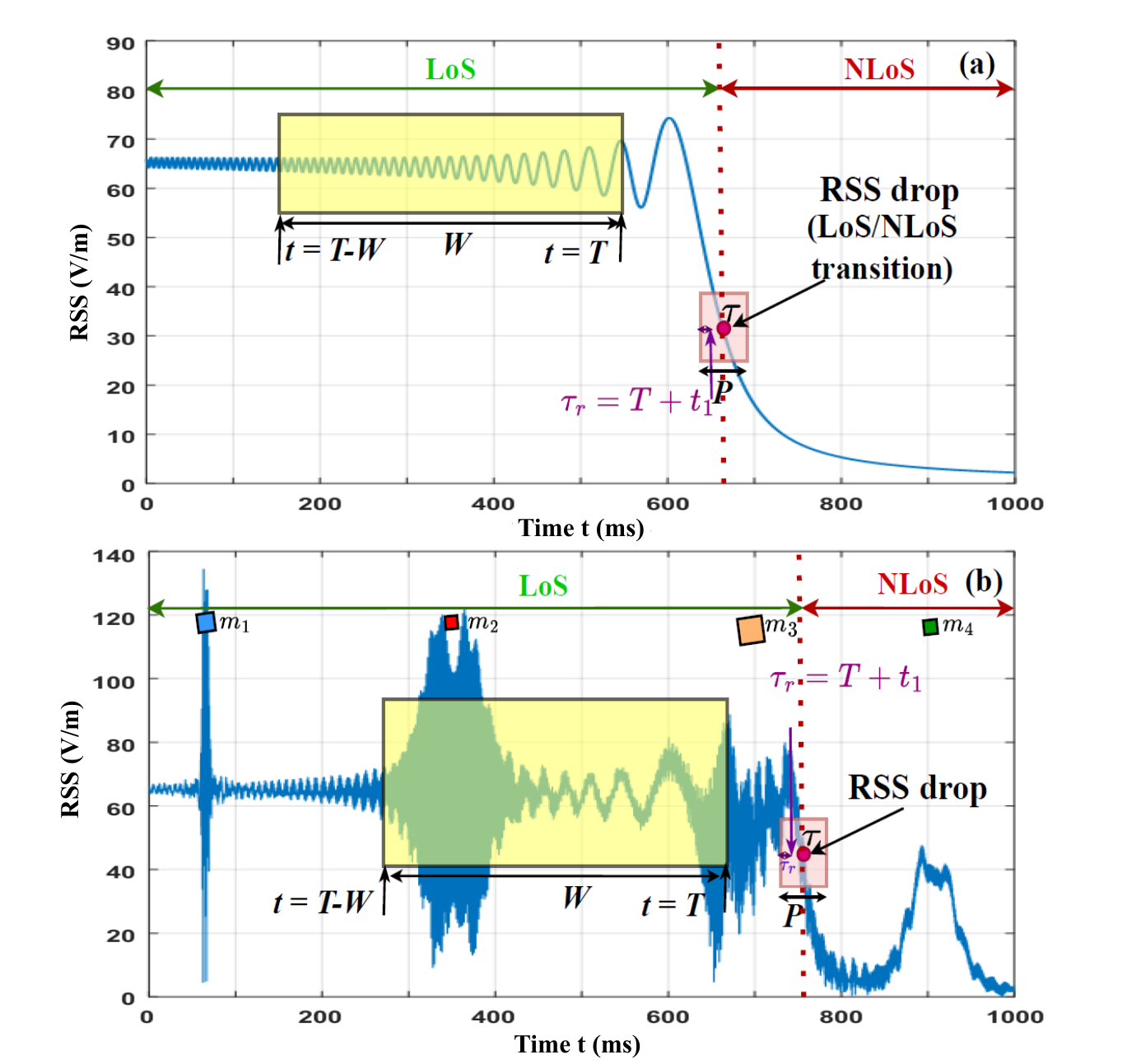}}
  \caption{RSS ${|}h'_{k}(t){|}$ (\ref{eq3}) along the UE path from (${x_{0}}$,${y_{0}}$) to (${x_{1}}$,${y_{1}}$) in Fig. \ref{fig1}. (a) without reflectors, $Q = 0$; (b) with $Q = 4$ stationary reflectors; sizes and angles (counter-clockwise (CCW) from x-axis): $m_{1}$: 0.98 $m$, 95${^\circ}$; $m_{2}$: 0.7 $m$, 95${^\circ}$; $m_{3}$: 1.5 $m$, 98${^\circ}$; $m_{4}$: 0.78 $m$, 96.4${^\circ}$. Examples of the observation window of length $W$, prediction window of length $P$, end of observation window $T$, prediction range $t_1$, RSS drop time $\tau$, random time instance within prediction window $\tau_r$; $f_{k}$ = 30 GHz, UE speed $v$ = 8 $m/s$, MIMO $L = 16$ and $M =  4$.}\label{fig2}
\end{figure}

In Fig. \ref{fig2}, we illustrate the downlink ${|}h'_{k}(t){|}$ for the scenario of Fig. \ref{fig1}. Fig. \ref{fig2}(a) shows diffraction-induced oscillation patterns in the absence of reflectors in the environment. In the LoS region, preceding the geometric LoS/NLoS transition at $\tau$ = 680 $ms$, the oscillations tend to grow in amplitude and decrease in frequency. These patterns are calculated using Fresnel diffraction in the physical model \cite{b8}. As discussed i oscillation patterns occur tens of 5G frames prior to  the RSS drop caused by blockage for highway speeds \cite{b8}. 

Fig. \ref{fig2}(b) depicts the RSS ${|}h'_{k}(t){|}$ (\ref{eq3}) affected by the 4 reflectors shown in Fig. \ref{fig1}, illustrating both multipath fading and diffraction-induced oscillations. However, the former have higher frequency than the latter \cite{b8}. Due to narrow antenna beams, only one reflector at a time affects the RSS (\ref{eq3}). Thus, multipath fading occurs over the region in which both the reflector and the LoS path are present in the antenna lobes, such as a short reflection $m_1$ near 80 $ms$, a longer reflection $m_2$ from 270-420 $ms$, and $m_3$, which actively reflects into the interval 640-760 $ms$ where the LoS/NLoS transition occurs, thus extending the RSS drop $\tau$. Moreover, in the 890-930 $ms$ interval, the reflection $m_4$ occurs in the NLoS region, dominating the multipath signal created by its interference with the remnant, diffracted LoS near the transition. Finally, diffraction-induced oscillations occur when the reflection starts and prior to its end (just inside the reflector edges) \cite{b7,b18}. In summary, the above reflector effects and antenna models were not included in \cite{b8} and provide more realistic characterization of mmWave/sub-THz channels necessary for development of accurate ML-EW methods.

\section{ML-EW of BLOCKAGE for REALISTIC MIMO SYSTEMS}
 In this paper, we use the MiniRocket classifier \cite{b4} to extract features from the time-correlated samples of ${|}h'_{k}(t){|}$ (\ref{eq3}).  MiniRocket distinguishes itself from deep-learning models, such as CNN, RNN and Long Short-Term Memory (LSTM) networks, by employing a fixed set of convolutional kernels with diverse dilation and padding configurations and extracting features through a lightweight, deterministic process. A comprehensive evaluation in \cite{b17} demonstrates that Rocket-based methods not only achieve competitive accuracy but also outperform state-of-the-art models, such as TapNet, in terms of computational efficiency.
 
 We employ MiniRocket in the ML-EW method to identify presence or absence of blockage along the UE trajectory within the prediction window based on the RSS samples within the observation window, where the two windows are separated by the prediction interval. MiniRocket employs 10,000 convolutional kernels of length 12 with random weights and dilation to detect the pre-blockage diffraction-induced oscillations within the observation window \cite{b8}. It utilizes a ridge regression classifier—a linear least-squares method with $\ell_2$ regularization—for efficient prediction \cite{b5}. This paper investigates the ML-EW algorithm's ability to distinguish the pre-blockage oscillations from realistic multipath fading with varying reflectors' sizes and angles for diverse MIMO scenarios and different carrier frequencies, which was not investigated in \cite{b8}.
 
 Fig. \ref{fig2} shows an example of the parameters that are fixed in each ML-EW algorithm implementation: the length of the observation window $W$ ($ms$), the length of the prediction interval (the prediction range) $t_{1}$ ($ms$), and the length of the prediction window $P$ ($ms$). The RSS is sampled at the rate $f_{s}$ (Hz). Moreover, Fig. \ref{fig1} depicts one example of a physical model setting used to generate the training and testing dataset $\mathcal{D}$ = \{${\mathcal{X}_{i}}$,${\mathcal{L}}_{i}$\} $i =1, . . . ,N$, where ${\mathcal{X}_{i}}$ is a 1 $ms$-long scenario with specified BS, reflectors, blocker, and UE positions, trajectories, speeds, sizes, angles, etc, and the label ${\mathcal{L}}_{i}$ indicates absence (${\mathcal{L}}_{i}$ = 0) or presence (${\mathcal{L}}_{i}$ = 1) of blockage within the UE trajectory of this ${\mathcal{X}_{i}}$.

When ${\mathcal{L}}_{i} = 1$, blockage occurs after $W + t_1$ ($ms$) of the ${\mathcal{X}_{i}}$ trajectory. In most of these scenarios, the RSS drop $\tau$ is due to the geometric LoS/NLoS transition, e.g., 680 ($ms$) in Fig. \ref{fig2}(a), which occurs at 50\% RSS value of the LoS RSS \cite{b7}. However, occasionally a reflector overlaps with the LoS/NLoS transition, e.g., $m_3$ in Fig. \ref{fig2}(b), and extends the time instance of the RSS drop $\tau$, after which the RSS falls below 50\% of the initial LoS RSS value for this UE trajectory. Once $\tau$ is determined for a given trajectory, we position this time instance $\tau$ as the midpoint of the prediction window, which is $P$ ($ms$) long (see Fig. \ref{fig2}). When ${\mathcal{L}}_{i}$ = 0, the location of the prediction window is chosen randomly using the uniform distribution after the last $W + t_1$ ($ms$) of the UE trajectory. After the prediction window is fixed, we randomly choose a point $\tau_{r}$ ($ms$) within the prediction window and place the end of the observation window of length $W$ ($ms$) at $T = \tau_{r} - t_{1}$. 

To test the effectiveness of the ML-based early warning method, we use k-fold cross-validation with $k$ = 5. The dataset is randomly split into 5 groups, $\mathcal{D}_{1}$ to $\mathcal{D}_{5}$. The model is trained on four groups at a time and tested on the remaining group to evaluate its prediction performance. This process is repeated five times, so that each group is tested once. During training, the RSS (\ref{eq3}) of each UE trajectory $\mathcal{X}_{i}$ in the training subset is sampled at the rate $f_{s}$ (Hz) within the observation window of length $W$ ($ms$). The features of these samples, extracted using MiniRocket, are associated with the predetermined label $\mathcal{L}_{i}$. Note that random choice of $\tau_r$ within the prediction window facilitates variation of the observation window position, thus improving learning and prediction accuracy. This impact is enhanced by increasing the length $P$ of the prediction window as shown in section IV.

When testing, the algorithm observes the samples of a mobile trajectory in the testing group within the window of length $W$ ($ms$) (positioned as explained above for datasets with and without blockage) and classifies the corresponding dataset as "blockage" ${\mathcal{\hat{L}}}_{i}$ = 1 or "absence of blockage" ${\mathcal{\hat{L}}}_{i}$ = 0. If ${\mathcal{L}}_{i} = {\mathcal{\hat{L}}}_{i}$, the dataset $\mathcal{X}_{i}$ is classified correctly.

The UE and blocker trajectories for each scenario are contained in the rectangular region formed by $R_{1}$ = (0,0) $m$ and $R_{2}$ = (160,60) $m$ (see Fig. \ref{fig1}). The UE trajectory originates from a random point in LoS and terminates in either LoS or NLoS to generate a rich dataset of various scenarios. The blocker starts its movement from point B = (0,0) $m$. In each scenario, BS is stationary and is placed randomly along the line between (-40,0) $m$ and (-0.5,0) $m$. We create a diverse dataset of scenarios by using independent, uniform distributions to select the directions and speeds of the UE and blocker between 0 and 30 $m/s$ as well as the distances and angles between the UE, the blocker and the BS positions. These variations aim to capture various diffraction-induced patterns and multipath with different rates and amplitudes. The dataset $\mathcal{D}$ is balanced, with 50\% of the scenarios labeled as $\mathcal{L}_{i}$ = 1 and the other 50\% labeled as $\mathcal{L}_{i}$ = 0. This ensures that the model trained on this dataset will not be biased towards one class.  

Moreover, BS and UE are equipped with uniform linear dipole antennas array of sizes $L$ and $M$, respectively, as described in section II, and variety of sizes, orientations and speeds of the reflectors are employed. The number of dominant reflectors is expected to be small in 5G/6G systems due to utilization of MIMO \cite{b3}. In this paper, we only consider reflectors with flat reflecting surfaces as {the contribution of a curved reflector quickly weakens with distance from the reflector  \cite{b7,b18}. Thus, we define the size of the reflector as its length in 2D, with the position coordinates at the center and its angle measured CCW from the positive x-axis (see Fig. \ref{fig1}). To generate each scenario, we position $Q$ significant reflectors randomly using the uniform distribution, where $Q$ is chosen using equiprobable distribution in the set from 0 and 10. Reflector sizes range from 0.1 $m$ to 2 $m$ while reflectors’ angles are randomized using the uniform distribution between 0${^\circ}$ and 360${^\circ}$. For scenarios with $Q \geq 8$, we randomly pick moving/rotating reflectors as follows. When $Q = 8$, one reflector moves or rotates. For $Q = 9$, there are two moving or two rotating reflectors or one moving and one rotating reflector. For $Q = 10$, two reflectors move and one rotates or one reflector moves and two rotate. Equiprobable distributions are used for all outcomes above.  Reflectors move with speeds that vary randomly with the uniform distribution between 0 and 30 $m/s$ or rotate with randomly and uniformly distributed initial and final angles between 0${^\circ}$ and 360${^\circ}$ from positive x-axis (see Fig. \ref{fig1}).   

\begin{table}[]
\begin{minipage}[t]{0.45\textwidth}
    \begin{center}
            \caption{ML-EW PERFORMANCE vs DATASET SIZE $N$, $W$ = 400$ms$, $f_{s}$ = 1kHz, $t_{1}$ = 100$ms$, $P$ = 50$ms$, $f_{k}$ = 30GHz, $L$ = 16, $M$ = 4, $Q \leq 10$.}
            \resizebox{5.3cm}{!}{
            \small
                \begin{tabular}{|c|c|c|c|}
                \hline
                \textit{N} & Accuracy & F1 score & AUC   \\ \hline
                4000       & 89.87\%  & 89.86\%    & 89.87\% \\ \hline
                6000       & 90.30\%  & 90.29\%    & 90.28\% \\ \hline
                8000       & 90.52\%  & 90.51\%    & 90.52\% \\ \hline
                10000      & 90.77\%  & 90.77\%    & 90.76\% \\ \hline
                12000      & 90.75\%  & 90.75\%    & 90.74\% \\ \hline
                14000      & 90.77\%  & 90.76\%    & 90.76\% \\ \hline
                15000      & 90.80\%  & 90.81\%    & 90.80\% \\ \hline
            \end{tabular}
            \medskip}
            \label{Tab:Tab1}
    \end{center}
\end{minipage}
\end{table}

\begin{table}[]
\begin{minipage}[t]{0.45\textwidth}
    \begin{center}
            \caption{ML-EW PERFORMANCE vs SAMPLING FREQUENCY $f_{s}$, 
              \\$N$ = 10000, $W$ = 400$ms$, $t_{1}$ = 100$ms$, $P$ = 50$ms$, $f_{k}$ = 30GHz, \\$L$ = 16, $M$ = 4, $Q \leq 10$.}
            \resizebox{5.3cm}{!}{
            \small
                \begin{tabular}{|c|c|c|c|}
                \hline
                \textit{fs} (Hz) & Accuracy & F1 score & AUC   \\ \hline
                1000        & 90.77\%  & 90.76\%    & 90.74\% \\ \hline
                2000        & 91.39\%  & 91.38\%    & 91.38\% \\ \hline
                3000        & 91.63\%  & 91.62\%    & 91.62\% \\ \hline
                4000        & 92.32\%  & 92.31\%    & 92.31\% \\ \hline
                5000        & 92.24\%  & 92.23\%    & 92.23\% \\ \hline
            \end{tabular}
            \medskip}
            \label{Tab:Tab2}
    \end{center}
\end{minipage}
\end{table}

To evaluate the performance of our ML-EW method, we use three metrics: accuracy (the percentage of correctly predicted labels), $f_{1}$ score, and area under the curve (AUC) \cite{b5}. 

 \begin{table}[]
\begin{minipage}[t]{0.45\textwidth}
    \begin{center}
            \caption{ML-EW PERFORMANCE vs PREDICTION RANGE $t_{1}$ and INTERVAL $P$, $W$ = 400$ms$, $f_{s}$ = 4kHz, $M$ = 4, $L$ = 16, \\$f_{k}$ = 30GHz, $Q \leq 10$.}
            \resizebox{6cm}{!}{
                \begin{tabular}{ccccc}
                \hline
                \textit{P}(ms) & \textit{$t_{1}$}(ms)           & Accuracy & F1 score & AUC   \\ \hline
                10    & 100 & 91.86\%  & 91.85\%    & 91.85\% \\
                10    & 150              & 90.05\%  & 90.04\%    & 90.04\% \\
                10    & 200              & 88.29\%  & 88.27\%    & 88.27\% \\
                10    & 250              & 85.67\%  & 85.65\%    & 85.67\% \\
                10    & 350              & 81.62\%  & 81.61\%    & 81.61\% \\ \hline
                25    & 100  & 92.22\%  & 92.21\%    & 92.21\% \\
                25    & 150              & 90.78\%  & 90.77\%    & 90.76\% \\
                25    & 200              & 88.34\%  & 88.32\%    & 88.32\% \\
                25    & 250              & 85.74\%  & 85.71\%    & 85.71\% \\
                25    & 350              & 81.71\%  & 81.70\%    & 81.70\% \\ \hline
                50    & 100  & 92.32\%  & 92.31\%    & 92.31\% \\
                50    & 150              & 90.90\%  & 90.89\%    & 90.89\% \\
                50    & 200              & 88.72\%  & 88.73\%    & 88.70\% \\
                50    & 250              & 85.81\%  & 85.80\%    & 85.80\% \\
                50    & 350              & 81.90\%  & 81.90\%    & 81.89\% \\ \hline
            \end{tabular}}
            \label{Tab:Tab3}
    \end{center}
\end{minipage}
\end{table}

\begin{table}[]
\begin{minipage}[t]{0.5\textwidth}
    \begin{center}
            \caption{ML-EW ACCURACY for MIMO ($L$ = 16, $M$ = 4) and  OMNIDIRECTIONAL ANTENNAS,  $f_{s}$ = 4kHz, $W$ = 400$ms$,\\ $P$ = 50$ms$, $f_{k}$ = 30GHz.}
            \resizebox{6.7cm}{!}{\begin{tabular}{|p{1.3cm}|p{1.3cm}|p{1.3cm}|p{2.3cm}|}
            \hline
            \textit{$t_{1}$}(ms) & MIMO \hfil ($Q$=0)& MIMO \hfil($Q\leq$10) & Omnidirectional \hfil ($Q\leq$10) \\ \hline
            \hfil 100 & \hfil 97.93\% & \hfil 92.32\% & \hfil 88.11\%         \\ \hline
            \hfil 150              & \hfil 97.38\% & \hfil 90.89\% & \hfil 85.73\%         \\ \hline
            \hfil 200              & \hfil 96.55\% & \hfil 87.73\% & \hfil 83.52\%         \\ \hline
            \hfil 250              & \hfil 95.50\% & \hfil 85.67\% & \hfil 80.86\%         \\ \hline
            \hfil 350              & \hfil 94\%    & \hfil 81.90\% & \hfil 77.90\%         \\ \hline
            \end{tabular}}
    \label{Tab:Tab4}
    \end{center}
\end{minipage}
\end{table}

\section{NUMERICAL RESULTS}

This section evaluates the effectiveness and accuracy of the ML-EW method for MIMO mmWave/sub-THz channels with reflectors in the environment.

\subsection{EW Performance}

In this subsection, we fix $f_{k}$ = 30 GHz and $L = 16$, $M = 4$. The effects of varying these parameters will be examined in the following subsections. In Tables \ref{Tab:Tab1} and \ref{Tab:Tab2}, we compare the accuracy of ML for a varying dataset sizes $N$  and sampling frequencies $f_{s}$. As in \cite{b8}, we found that choosing the observation window length $W$ = 400 $ms$ provides desirable performance complexity trade-off. From Table \ref{Tab:Tab1}, we note that increasing $N$ to values above 10000 does not significantly improve performance. Moreover, from Table \ref{Tab:Tab2}, ML performance improves as $f_{s}$ increases and saturates at $f_{s}$ = 4 kHz. Similar conclusions were reached for other MIMO configurations and carrier frequencies. Thus, we choose $N$ = 10000, $W$ = 400 $ms$ and $f_{s}$ = 4 kHz in the remainder of this paper. Table \ref{Tab:Tab3} shows that performance of ML-EW improves as the prediction time $t_{1}$ decreases and the prediction window length $P$ increases as expected. These results demonstrate that the ML-EW method achieves accuracy of over 90\% for $t_{1}$ = 150 $ms$ in the reflector-rich environments, corresponding to 15 frames or hundreds of slots in 5G networks. This prediction range is sufficient for a proactive response to the upcoming blockage.  

\begin{figure}[]
\centerline{
\includegraphics[width=0.47\textwidth, height=12cm]{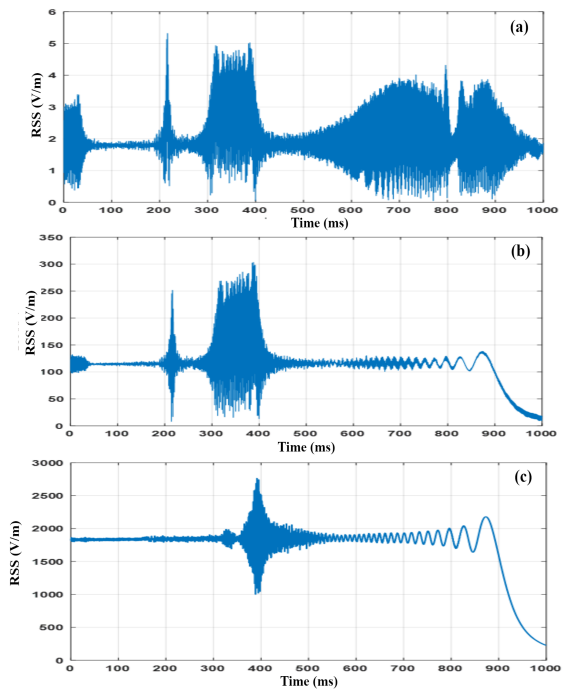}}
  \caption{RSS ${|}h'_{k}(t){|}$ (\ref{eq3}) for the same set of 9 reflectors (a) omnidirectional antennas; (b) MIMO with $M$ = 4, $L$ = 16; (c) MIMO with $M$ = 16, $L$ = 64, $f_{k}$ = 30GHz; UE speed = 17$m/s$, blocker speed = 9$m/s$.}\label{fig:fig3}
\end{figure}

\subsection{Impact of MIMO and Reflectors}

In Fig. \ref{fig:fig3}, we compare the RSS of MIMO with that of the omnidirectional antennas for a trajectory with Los/NLoS transition and 9 reflectors, where one of the reflectors is moving with the speed 2 $m/s$ and one is rotating with the initial/final angles given by 97${^\circ}$ and 320${^\circ}$, respectively. Due to the narrower beamwidths of MIMO signals, the impact of reflectors on the RSS is smaller than for omnidirectional antennas.

For each antenna array configuration, the ML-EW algorithm is trained and tested separately. Table \ref{Tab:Tab4} shows its performance for MIMO and omnidirectional antennas with and without reflectors in the environment. We found that performance is similar for MIMO and omnidirectional antennas {when $Q = 0$, i.e.,  in the absence of reflectors. However, presence of reflectors degrades ML performance due to the additional oscillation patterns and RSS fluctuations as discussed in section II. Furthermore, reflection around the LoS/NLoS transition (Fig. \ref{fig1}) can extend the RSS drop as shown in Fig. 2(b) but does not affect the RSS within the observation window and cannot be predicted in advance, thus further reducing the accuracy of ML-EW. However, employing MIMO improves the performance of EW up to 5\% in the environments with reflectors ($Q \leq 10$). ML-EW results for different array sizes are presented in Table \ref{Tab:Tab5}.  With $L$ = 64 and $M$ = 16, ML-EW achieves the accuracy of 96\%, an improvement of about 4\% compared to $L$ = 16 and $M$ = 4. This gain can be explained by the fact that the number of significant reflectors within the mobile trajectory decreases as the antenna size grows, as shown in Fig. \ref{fig:fig3}.

\begin{table}[]
\begin{minipage}[t]{0.45\textwidth}
    \begin{center}
            \caption{ML-EW PERFORMANCE vs ARRAY SIZES $L$ and $M$; \\$f_{s}$ = 4KHz, $W$ = 400$ms$, $t_{1}$ = 100$ms$, $P$ = 50$ms$ and \\$f_{k}$ = 30GHz, $Q \leq 10$.}
            \resizebox{6cm}{!}{
            \small
                \begin{tabular}{|l|l|l|l|}
                \hline
                \multicolumn{1}{|l|}{Array size} & Accuracy & F1 score & AUC   \\ \hline
                \textit{L} = 16,  \textit{M} = 4                      & 92.32\%  & 92.31\%    & 92.31\% \\ \hline
                \textit{L} = 32, \textit{M} = 4                       & 93.07\%  & 93.06\%    & 93.06\% \\ \hline
                \textit{L} = 48, \textit{M} = 4                       & 93.70\%  & 93.69\%    & 93.69\% \\ \hline
                \textit{L} = 16, \textit{M} = 8                       & 93.39\%  & 93.38\%    & 93.38\% \\ \hline
                \textit{L} = 16, \textit{M} = 16                      & 94.43\%  & 94.42\%    & 94.42\% \\ \hline
                \textit{L} = 64, \textit{M} = 16                      & 96.06\%  & 96.05\%    & 96.05\% \\ \hline
            \end{tabular}
             \medskip}
            \label{Tab:Tab5}
    \end{center}
\end{minipage}
\end{table}

\subsection{Dependency on the Carrier Frequency}

Table \ref{Tab:Tab6} presents the performance of ML-EW vs the carrier frequency $f_{k}$ in the mmWave and sub-THz range. For each frequency $f_{k}$, the model is trained and tested separately. We observe that the accuracy of ML-EW is above 90\% for all mmWave/sub-THz frequencies, with slight drop in accuracy as $f_{k}$ increases, consistently with the derivations in \cite{b8} that show the diffraction-induced oscillation patterns preceding blockage retain their features but compress in space by a factor of $\sqrt{f_{k}}$. The effect of $f_{k}$ on the oscillation patterns prior to blockage with and without reflectors is depicted in Fig. 4.

\begin{figure}[]
\centerline{
\includegraphics[width=0.47\textwidth, height=12.5cm]{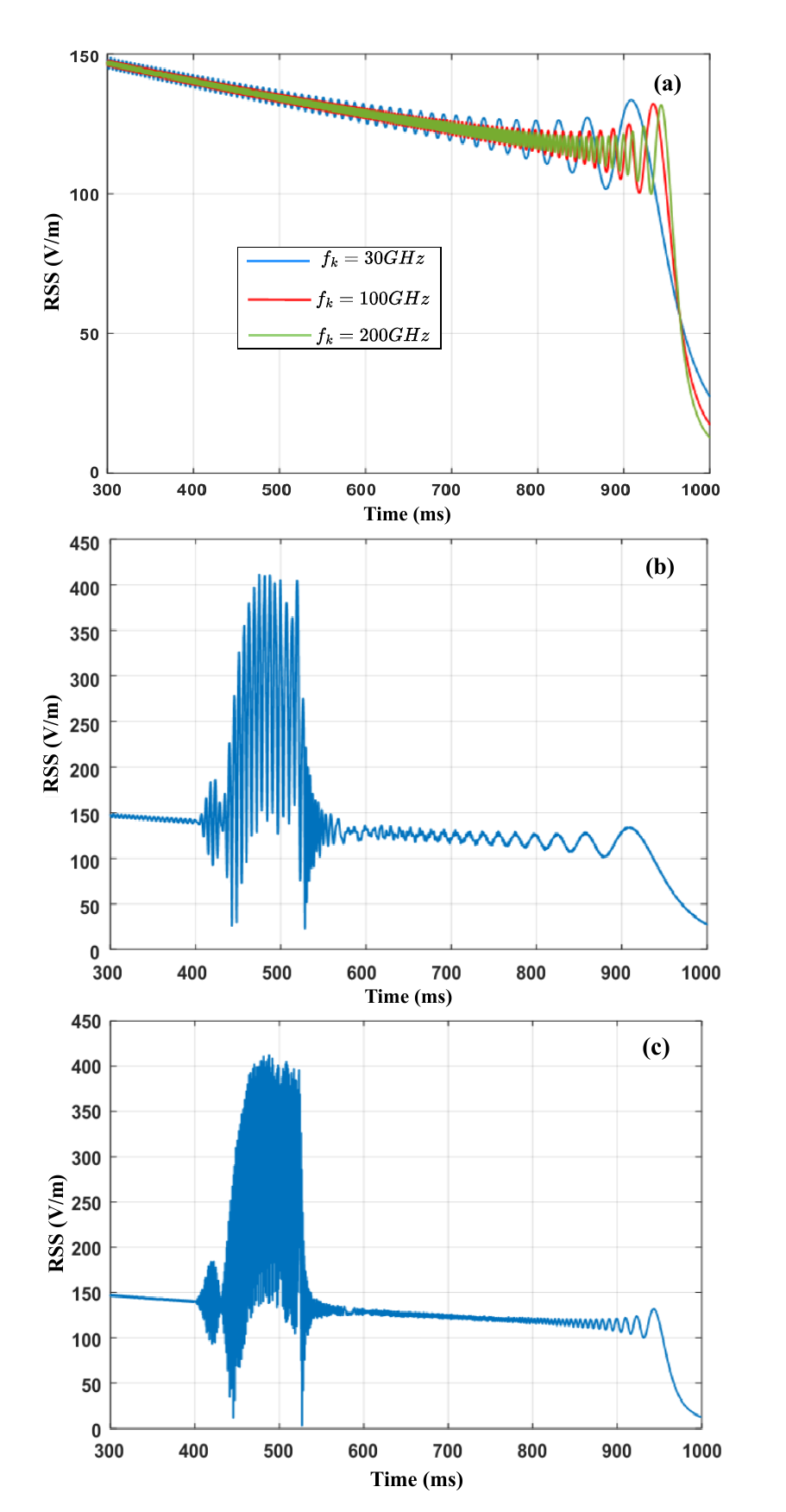}}
  \caption{RSS ${|}h'_{k}(t){|}$ (\ref{eq3}) with $M$ = 4, $L$ = 16; (a) in the absence of reflectors ($Q$ = 0); (b) $f_{k}$ = 30GHz and (c) $f_{k}$ = 200GHz with $Q$ = 2 stationary reflectors; UE speed = 23$m/s$, blocker speed = 21$m/s$.}\label{fig:fig4}
\end{figure}

\begin{table}[]
\begin{minipage}[t]{0.45\textwidth}
    \begin{center}
            \caption{EW PERFORMANCE vs CARRIER FREQUENCY $f_{k}$, $L$ = 16, \\$M$ = 4, $f_{s}$ = 4KHz, $W$ = 400$ms$, $t_{1}$ = 100$ms$, $P$ = 50$ms$, $Q \leq 10$.}
            \resizebox{5.7cm}{!}{
            \small
                \begin{tabular}{|c|c|c|c|}
                \hline
                \multicolumn{1}{|l|}{$f_{k}$ (GHz)} & Accuracy & F1 score & AUC   \\ \hline
                30                             & 92.32\%  & 92.31\%    & 92.31\% \\ \hline
                60                             & 91.96\%  & 91.95\%    & 91.95\% \\ \hline
                80                             & 91.75\%  & 91.74\%    & 91.74\% \\ \hline
                100                      & 91.43\%  & 91.42\%    & 91.42\% \\ \hline
                150                      & 91.21\%  & 91.20\%    & 91.20\% \\ \hline
                200                      & 90.58\%  & 90.57\%    & 90.57\% \\ \hline
                250                      & 90.44\%  & 90.43\%    & 90.43\% \\ \hline
                300                      & 90.32\%  & 90.31\%    & 90.30\% \\ \hline
            \end{tabular}
            \medskip}
            \label{Tab:Tab6}
    \end{center}
\end{minipage}
\end{table}

\subsection{Impact of Mobility}

The effect of UE, blocker and reflector speeds on the EW performance is shown in the Table \ref{Tab:Tab7}.  In this subsection, the number of reflectors is $2 \leq Q \leq 10$. There are two moving reflectors for all the $Q$ values, and for $Q = 9,10$, there is also one rotating reflector, where all outcomes are chosen according to an equiprobable distribution. Training of ML is performed jointly for all speeds for each iteration $k$ of the $k$-fold cross-validation method using the original dataset $\mathcal{D}$ (see section III). A randomly generated, balanced testing dataset of 2000 scenarios is created separately in each iteration $k$ for each of the 9 speed subsets in the Table \ref{Tab:Tab7}. Within a fixed UE/blocker speed subset, we observe that the effect of reflectors on RSS decreases slightly as the reflector speed increases and, therefore, the accuracy of ML-EW increases, which can be explained by shorter durations and higher oscillation rates of faster reflectors. On the other hand, within each reflector speed subset, the accuracy of ML-EW decreases slightly as UE/blocker speeds increase since pre-blockage oscillation patterns compress in time as mobility increases \cite{b8}. However, our results show that EW of blockage is still feasible for highway speeds.

\begin{table}[]
\begin{minipage}[t]{0.43\textwidth}
    \begin{center}
            \caption{ML-EW ACCURACY for MIMO SYSTEM ($L$ =16, $M$ =4) vs UE and BLOCKER ($v_{UE,B}$) and REFLECTOR ($v_{q}$) SPEEDS, $f_{s}$=4KHz, $W$ = 400$ms$, $t_1 = 100$$ms$, $P$ = 50$ms$, \\$f_{k}$ = 30GHz, $2 \leq Q \leq 10$.}

           \resizebox{5cm}{!}{
            \begin{tabular}{|@{}l|c|c|c|c|r@{}|}
            \hline
            \diagbox[width=4em,trim=l]{\small $v_{\tiny UE,\small{B}}$}{$v_{q}$} & 0-10 & 10-20 & 20-30 \\
            \hline
             \hspace{2mm}  0-10 & 89.01\%    &   90.02\%   & 90.17\% \\
            \hline
             \hspace{1.2mm} 10-20 & 88.11\%    &   89.22\%   & 90.08\% \\
            \hline
             \hspace{1.2mm} 20-30 & 87.94\%      &   88.21\%   & 89.50\% \\
             \hline        
            \end{tabular}
             \medskip}
    \label{Tab:Tab7}
    \end{center}
\end{minipage}
\end{table}

\section{Conclusion}

In this work, we developed and evaluated the performance of an ML-based early warning of LoS blockage method using in-band observations for mmWave and sub-THz systems in environments with reflectors and directional antennas at the BS and the UE. The proposed ML-EW method is trained and tested using our realistic physical model that accurately models the multipath and diffraction oscillation patterns associated with reflectors and approaching blockers and facilitates performance evaluation over diverse topologies, frequencies, speeds, angles, reflector, and MIMO configurations. We found that presence of reflectors and higher carrier frequencies degrade prediction accuracy while additional antennas enhance it. Overall, the proposed EW method forecasts approaching blockage at least 100 $(ms)$ ahead with accuracy about or exceeding 90\% in realistic mobile 5G/6G systems, thus enabling a proactive response, such as finding a new beam, performing a handover between base stations, or updating resource allocation.

\vspace{12pt}
\color{red}


\begin{thebibliography}{00}
\bibitem{b2} T. S. Rappaport, Y. Xing, O. Kanhere, S. Ju, A. Madanayake, S. Mandal, A. Alkhateeb, and G. C. Trichopoulos, “Wireless communications and applications above 100 ghz: Opportunities and challenges
for 6g and beyond,” IEEE Access, vol. 7, pp. 78 729–78 757, 2019.
\bibitem{b3} T. S. Rappaport, G. R. MacCartney, M. K. Samimi, and S. Sun, “Wideband millimeter-wave propagation measurements and channel models for future wireless communication system design,” IEEE Transactions on Communications, vol. 63, no. 9, pp. 3029–3056, 2015.
\bibitem{b4} A. Dempster, D. F. Schmidt, and G. I. Webb, “Minirocket: A very fast (almost) deterministic transform for time series classification,” in Proceedings of the 27th ACM SIGKDD Conference on Knowledge Discovery ${\&}$ Data Mining, 2021, pp. 248–257.
\bibitem{b5} C. M. Bishop and N. M. Nasrabadi, Pattern Recognition and Machine Learning, vol. 4, no. 4. New York, NY, USA: Springer, 2006.
\bibitem{b7} Z. Ali, A. Duel-Hallen, and H. Hallen, “Early warning of mmWave signal blockage and AoA transition using sub-6 GHz observations,” IEEE Communications Letters, vol. 24, no. 1, pp. 207–211, 2019.
\bibitem{b8} A. Fallah Dizche, A. Duel-Hallen and H. Hallen, "Early Warning of mmWave Signal Blockage Using Diffraction Properties and Machine Learning," in IEEE Communications Letters, vol. 26, no. 12, pp. 2944-2948, 2022.
\bibitem{b9} L. Yu, J. Zhang, Y. Zhang, X. Li and G. Liu, "Long-Range Blockage Prediction Based on Diffraction Fringe Characteristics for mmWave Communications," in IEEE Communications Letters, vol. 26, no. 7, pp. 1683-1687, July 2022.
\bibitem{b12} S. Wu, M. Alrabeiah, C. Chakrabarti and A. Alkhateeb, "Blockage Prediction Using Wireless Signatures: Deep Learning Enables Real-World Demonstration," in IEEE Open Journal of the Communications Society, vol. 3, pp. 776-796, 2022. 
\bibitem{b15} R. Heath, A. Lozano, \textit{Foundations of MIMO Communication}. Cambridge: Cambridge University Press, 2018, doi:10.1017/9781139049276.
\bibitem{b17} A. P. Ruiz, M. Flynn, J. Large, M. Middlehurst, and A. Bagnall, “The great multivariate time series classification bake off: A review and experimental evaluation of recent algorithmic advances,”.Data Min.
Knowl. Discov., vol. 35, no. 2, pp. 401–449, Mar. 2021.
\bibitem{b18} Z. Ali, A. Duel‑Hallen, and H. Hallen, “Supplementary Material for ‘Early Warning of mmWave Signal Blockage and AoA Transition Using sub‑6 GHz Observations,’”,  IEEE DataPort, 2019, [Online].  Available: https://ieeexplore.ieee.org/ielx7/4234/8952820/8895815/supp1-2952602.pdf.
\bibitem{b19} A.Fallah Dizche, A. Duel‑Hallen, and H. Hallen,"Supplementary Material for Early Warning of mmwave Signal Blockage Using Diffraction Properties and Machine Learning", 2022 [Online]. Available:  https://github.com/amirdizche/mmWEW

\end{thebibliography}
\end{document}